\documentclass{revtex4-1}
\usepackage{mathrsfs}
\DeclareMathAlphabet{\mathcal}{OMS}{cmsy}{m}{n}
\usepackage{amsmath}
\usepackage{amssymb}
\usepackage{amsfonts}
\usepackage{array}
\usepackage{graphicx}

\begin{document}
{\centering {\huge Robust Digital Holography For Ultracold Atom Trapping}

\vspace{0.2cm}

Alexander L. Gaunt$^*$ and Zoran Hadzibabic

\vspace{0.2cm}

Cavendish Laboratory, J J Thomson Avenue, Cambridge, CB3 0HE, UK

$^*$alg51@cam.ac.uk

}

\begin{quote}
\textbf{Abstract} \hspace{0.5cm} We have formulated and experimentally demonstrated an improved algorithm for design of arbitrary two-dimensional holographic traps for ultracold atoms. Our method builds on the best previously available algorithm, MRAF, and improves on it in two ways. First, it allows for creation of holographic atom traps with a well defined background potential. Second, we experimentally show that for creating trapping potentials free of fringing artifacts it is important to go beyond the Fourier approximation in modelling light propagation. To this end, we incorporate full Helmholtz propagation into our calculations.
\end{quote}

\vspace{1cm}
\noindent Optical dipole traps have become ubiquitous in disciplines ranging from live cell manipulation in biophysics \cite{Ashkin87} to single atom manipulation for quantum information processing \cite{Beugnon07}. 
In the field of ultracold atomic gases a variety of optical potential shapes, of ever increasing complexity, are used for fundamental studies of many-body physics \cite{Bloch:2008} in different optical crystal lattices \cite{Greiner:2002a, Struck:2011}, reduced dimensionality \cite{Kinoshita:2004, Hadzibabic:2006}, and non-trivial trapping topologies \cite{Ramanathan:2011, Moulder:2012}. Optical sculpting is also likely to offer great benefits for trapped-atom interferometry with Bose-Einstein condensates (BECs) \cite{Shin:2004} and for creating ``atomtronic" \cite{Seaman07} optical circuits. 

The simplest optical trap for an ultracold atomic gas is formed by a single focused Gaussian laser beam \cite{StamperKurn:1998b}. Natural extensions include standing-wave optical lattice potentials produced by the interference of multiple laser beams \cite{Greiner:2002a}, and ring traps produced using Laguerre-Gauss laser modes \cite{Ramanathan:2011,Moulder:2012}. However there is so far no universal approach for creating an \textit{arbitrary} optical trapping potential on the required micrometer scale. Existing techniques fall into three categories: (1) a time-averaged potential can be created by the fast scanning of a focused laser beam \cite{Henderson09}, (2) an intensity mask or micromirror-device can be imaged onto the trapping plane of the atoms \cite{Scherer07, Liang09}, and (3) in the holographic method one manipulates the phase of the laser beam so as to create the desired intensity pattern after further propagation of the light. None of these methods is currently a clear overall winner; their relative merits depend in practice on various factors, such as the desired spatial resolution, efficiency of use of laser light, and temporal stability of the light pattern. 

In this paper, we provide a computational and experimental procedure for generating arbitrary potentials based on an improved holographic, phase only method. We illustrate our technique by producing high fidelity patterns which demonstrate suitable characteristics for atom trapping, opening a versatile toolbox for optical manipulation in exotic geometries.

\begin{figure}[b!]
\includegraphics[width=0.8\textwidth]{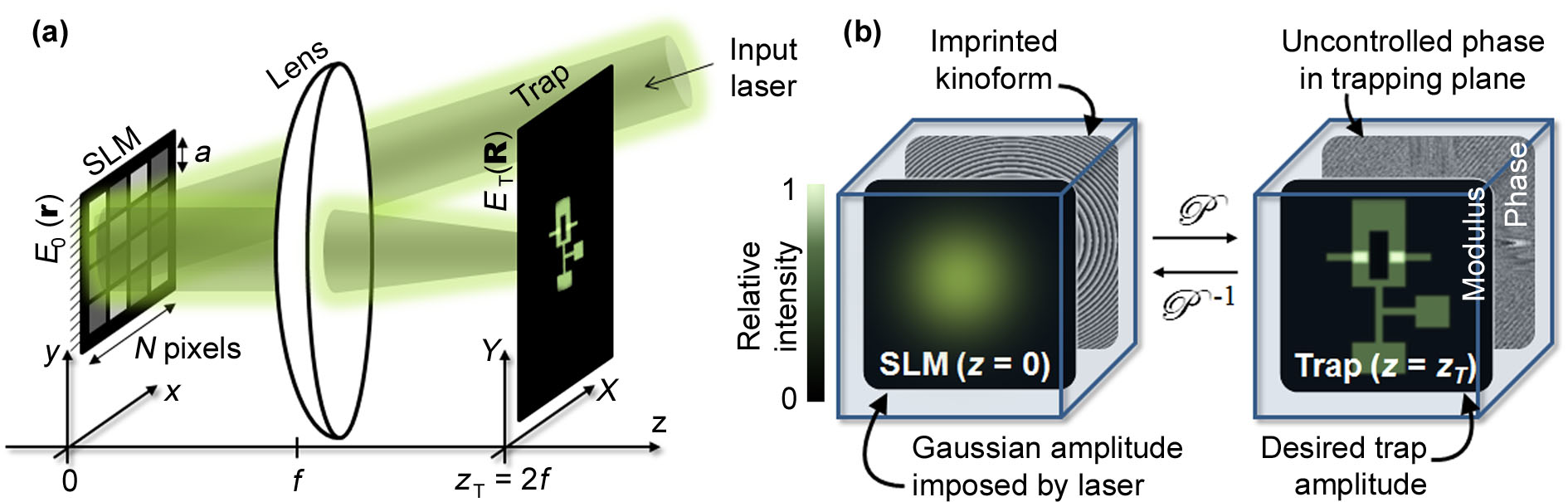}
\caption{ (a) Illustration of the optical system used for creating the traps. Collimated laser light is reflected from the SLM through a Fourier $2f$ arrangement to form the intensity pattern in the trapping plane. $f$ denotes the lens' focal length, see the Methods section for definition of the remaining symbols. The pattern shown in the trapping plane is the atomtronic OR-gate, which we use to test our method. (b) A cartoon of the constraints and the variables to be solved for. The modulus of the light field is fixed in the SLM plane, and we wish to solve for the kinoform to produce the correct modulus with no constraints on the phase in the trapping plane.\label{fig:fig1}}
\end{figure} 

As shown in Fig \ref{fig:fig1}(a), we use a spatial light modulator (SLM) to imprint a custom phase pattern (kinoform) on a laser beam in order to form a diffraction pattern of the desired shape in the trapping plane. Since this is a diffractive method, the kinoform steers the light into required trap shapes without blocking out any of the incident power, making the method potentially very efficient. In some cases, the required kinoform can be calculated analytically \cite{Ramanathan:2011, Moulder:2012}, but in general, this is a highly non-trivial numerical problem. The difficulty arises from the fact that we have incomplete \textit{a priori} knowledge of the light field in both planes: In the trapping plane only the \textit{modulus} of the light field is relevant for optical trapping, so we place no constraint on the phase; similarly, at the SLM, we impose the unaltered Gaussian \textit{modulus} of the incident laser profile, leaving the phase as the variable to be calculated [Fig. \ref{fig:fig1}(b)].

The numerical task of calculating the kinoform is approached as an optimisation procedure which can be achieved by ``steepest descent" methods such as direct binary searches \cite{Seldowitz87, Boyer04, Boyer06}, genetic algorithms \cite{Zhou99} or Gerchberg and Saxton's iterative Fourier transform methods \cite{Gerchberg72}. Our method is based on the Gerchberg-Saxton algorithm (see the Methods section) since it is very computationally efficient and is easily executed on a graphics processing unit. So far, the best computational results were obtained by Pasienski and DeMarco \cite{Pasienski08}, using an adaptation of the Gerchberg-Saxton scheme known as the Mixed-Region-Amplitude-Freedom (MRAF) algorithm. The idea of MRAF is to enhance convergence of the iterative algorithm in one region of the trapping plane by giving up control of the remaining regions \cite{Pasienski08,Wyrowski90,Akahori86,Aagedal96}. This approach produced excellent computational results for a range of generic trap shapes, and some shapes have also been successfully realised with laser light \cite{Bruce10, Bruce11}. However, one practical limitation of MRAF is that in order to achieve good convergence in bright areas of the trapping plane (green areas in Fig. \ref{fig:fig1}), we must surrender control over almost all of the dark, background regions. In particular, the original MRAF calculations provide for only a very narrow region defining the zero background potential to which all the potentials are referenced. This could lead to experimental problems such as inefficient loading of atoms into, or percolation of atoms out of the trap; this issue is particularly important if the desired trap has sharp edges so the region of uncontrolled light intensity is immediately adjacent to the region of high atomic density. If the region of zero potential is extended, the fidelity of MRAF kinoforms is severely compromised. The first result of this paper is to show that this problem can be eliminated by properly redefining the target potential so as to include both the conventionally defined trapping region and a sufficiently large well defined background region (a ``canvas" on which an arbitrary potential landscape is ``drawn"). We formulate an offset-MRAF (OMRAF) algorithm through a simple set of practical rules which obey the limits imposed by Nyquist's theorem and minimise the occurrence of convergence-stalling optical vortices.

The second result of this paper is to demonstrate that the paraxial approximation often employed in literature \cite{Pasienski08,Ersoy07} can lead to undesirable artifacts in the traps. We illustrate the improvement that can be made by direct numerical calculation of the light propagation using the Helmholtz equation. Using our computational algorithms and a simple apparatus, we experimentally demonstrate examples of optical traps suitable for atomic experiments, and compare the results to previous realisations \cite{Bruce10, Bruce11}.

The paper is organised as follows: First we compare computational results produced using the existing MRAF and our OMRAF algorithms. Then we present and analyse experimental images of light fields shaped by our method. Throughout this investigation, we primarily test our technique by creating an atomtronic OR-gate trap shape \cite{Pasienski08,Seaman07,Pepino10} (see Fig \ref{fig:fig1}). To illustrate the universality of our method we also include the final results for a uniform square trap and an annular BEC stirrer \cite{Pasienski08,Bruce10}.

\section*{Results}
\subsection*{The Gerchberg-Saxton algorithm and MRAF}
Within the Gerchberg-Saxton framework, the core algorithm relies on linking the moduli in the SLM and trapping planes by simulating the light propagation back and fourth between these planes. After each propagation, we manually impose the known modulus constraints while leaving the phase to converge on the required solution. The mathematical details are presented in the Methods section, and here we just briefly review the algorithms for kinoform calculation before presenting the numerical and experimental results.

It is well documented that convergence of the na\"ive Gerchberg-Saxton algorithm is highly erratic because the modulus constraints in each plane are non-convex \cite{Bauschke02, Levi84}. The MRAF algorithm improves on this by defining a ``drawing" $\mathcal{D}$ in the trapping plane containing all the points in which the desired potential is non-zero, and an additional ``canvas" region, $\mathcal{C}$, of zero potential around the drawing [Fig. \ref{fig:fig2}(a)]. The modulus constraint is then manually imposed only inside $\mathcal{C}$ and $\mathcal{D}$ at each iterative step, thus reducing the number of constraints the algorithm aims to satisfy.

In the original MRAF paper, region $\mathcal{C}$ is limited to a tight border around $\mathcal{D}$ which has a typical width of around 1\% of the trapping region length scales. Although $\mathcal{C}$ is a featureless region of zero potential, it is an integral part of the trap since it defines the zero to which all the pixels in $\mathcal{D}$ are referenced. For practical purposes such a small canvas can be a severe limitation. Our method aims to expand $\mathcal{C}$ to approach its theoretical limit which is set by the Nyquist theorem to be 25\% of the area of the simulated trapping plane (see the Methods section). In this large canvas regime, MRAF convergence stagnates and produces poor solutions even after many iterations (see Fig. \ref{fig:fig2})

\subsection*{Optical vortices and OMRAF}
\label{sec:opticalvorticesandomraf}
\begin{figure}
\centering
\includegraphics[width=1.0\textwidth]{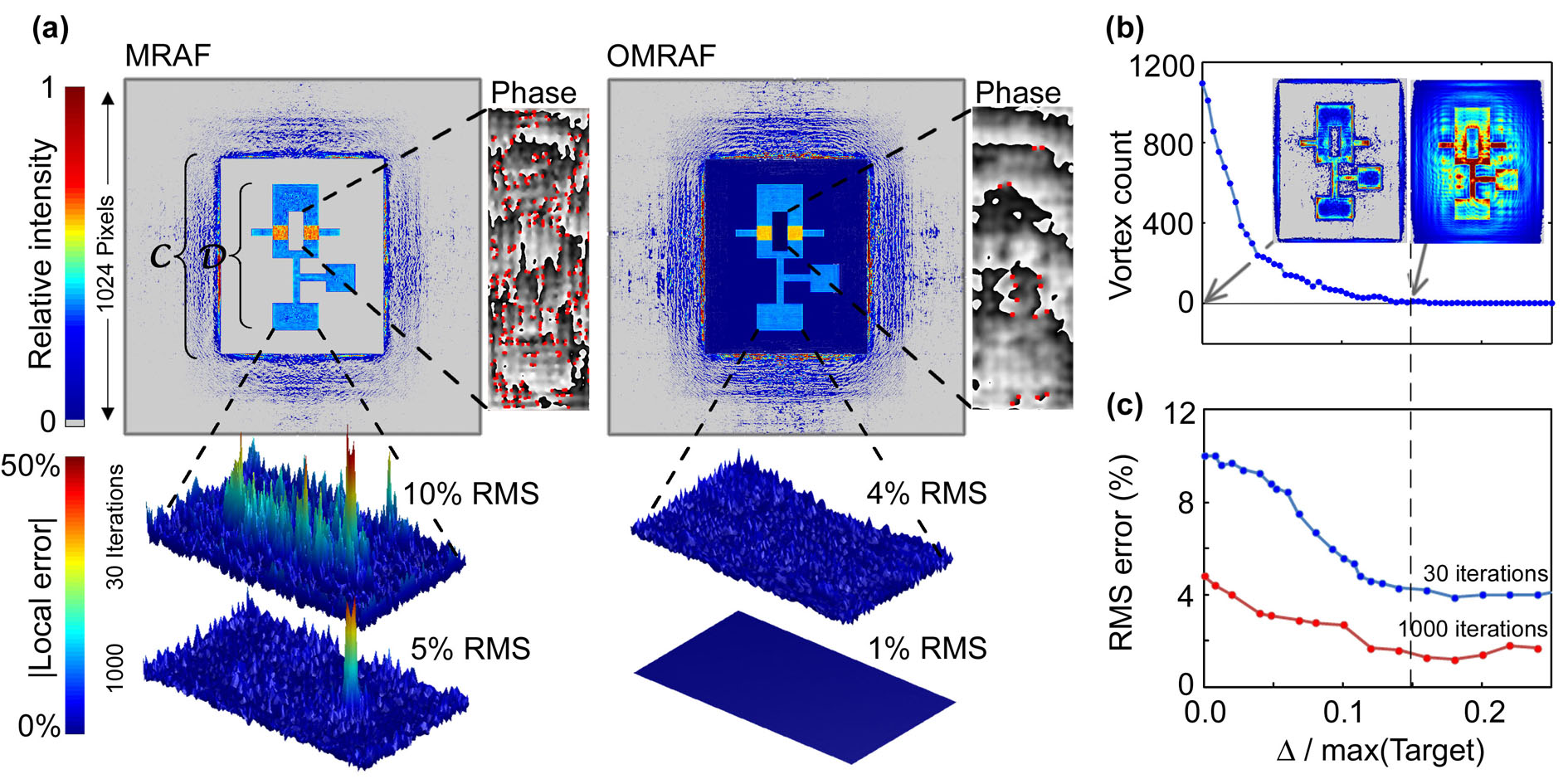}
\caption{Computational results for traps with a well defined zero potential background. (a) Reconstructions of the OR-gate trap with MRAF and OMRAF methods. We also show details of the local reconstruction error (bottom insets) and the local phase (right insets). In the phase plots, obtained after 30 iterations, we use black to white to represent 0 to $2\pi$ and red dots to mark vortex cores.  (b) Illustrates the method for picking the optimal offset, $\Delta$, measured in units of the peak amplitude in the target pattern. We perform a single iteration of the algorithm, for different offsets, to produce patterns such as those shown in the insets. We then choose the lowest value of $\Delta$ (indicated by the dashed line) which gives no vortices within $\mathcal{C}$ or $\mathcal{D}$ in this first iteration. (c) Improvement in the trap reconstruction associated with the offset.\label{fig:fig2}}
\end{figure}

We observe the problems of the MRAF algorithm for extended $\mathcal{C}$ to be due to the formation of large populations of optical vortices during the early iterations (approximately 1 vortex for every 10 pixels in Fig. \ref{fig:fig2}(a)). Optical vortices are topological features of the light field corresponding to a phase winding around a point of zero intensity. Large vortex populations are problematic because individual vortices are difficult to eliminate in further iterations of the algorithm. The reason for this is that the Gerchberg-Saxton algorithm monotonically reduces the RMS error in each iteration \cite{Fienup82}, but topological (un)winding operations cause disruption across the entire kinoform. Such global disruption is not in keeping with the monotonic error reduction in later iterations \cite{Wyrowski88}. Therefore any erroneous vortices are ``frozen in" early on, and the algorithm gets stuck in a local RMS minimum corresponding to a particular vortex distribution.

One possibility for handling vortices is by pairwise creation and annihilation, which requires only local operations; computational algorithms which encourage such pairwise operations are given in \cite{Senthilkumaran04}. However, we consider a much simpler solution. Our method simply offsets the trapping pattern inside $\mathcal{C}$ and $\mathcal{D}$ by a uniform intensity, $|\Delta|^2$, to remove all points of zero light intensity. This redefines the zero of the potential but does not change the physics of the trap. With an offset intensity, we know \textit{a priori} that we should aim to create no vorticies at all inside $\mathcal{C}$, so we feed the algorithm with an initial kinoform which contains no phase winding. Specifically, we choose a parabolic initial kinoform \cite{Aagedal96, Pasienski08} which defocuses the light into a patch with a characteristic size set by $\mathcal{C}$.

The optimal value of the offset intensity is calculated semi-empirically as follows: After just one iteration, the intensity in the trapping plane adopts approximately the correct shape, but exhibits considerable fluctuations around the desired intensity [Fig. \ref{fig:fig2}(b)]. If any of these fluctuations bring the intensity locally to zero, a vortex may form at this point. Therefore, we increase $|\Delta|^2$ until the trapping plane contains no vortices inside $\mathcal{C}$ after the first iteration [Fig. \ref{fig:fig2}(b) and (c)], and then trust that very few vortices will be formed in subsequent iterations. For maximum light-usage efficiency, we want to minimise the proportion of the incident light which is steered into the featureless background, i.e. choose the minimal $|\Delta|$ which gives satisfactory results. We find that for all the patterns we tested a suitable offset was $|\Delta| \sim 10-15\%$ of the maximum amplitude in the trapping plane (i.e. $|\Delta|^2\sim 1\%$ of the maximum intensity). The computational simulation of the trapping potential produced using the OMRAF algorithm is shown in Fig. \ref{fig:fig2}. Our method reduces the number of vortices seen in $\mathcal{C}$ after 30 iterations by a factor of $\sim20$ compared to MRAF. This allows excellent convergence of the algorithm, leaving only $4\%$ RMS error after 30 iterations and $1\%$ RMS after 1000.   

One compromise associated with the OMRAF method is a reduction in the efficiency of light use. We define efficiency as the ratio of the integral of the trap intensity above the background offset to the total integral across the trapping plane. Including the non-zero $\Delta$ causes a drop in efficiency by a factor of approximately 2, from 43\% to 24\%, for the OR-gate. Nevertheless, our method remains more efficient than intensity masking methods such as \cite{Scherer07}, which report 3\% efficiency for simple patterns. We calculate that for more complex patterns such as the OR-gate, intensity masking would be less than 1\% efficient.

\subsection*{Experimental Realisation with Laser Light}
We test our OMRAF algorithm experimentally by producing light patterns using 532 nm laser light in the arrangement shown in Fig. \ref{fig:fig1}(a). We use the standard SLM-based Shack-Hartmann algorithm \cite{Bowman10} to crudely correct for the low spatial frequency phase aberrations in the optical system and characterise our input laser beam. We then use an active feedback algorithm \cite{Bruce10} to optimise the final experimental patterns. This feedback routine is essential to remove the ``sinc envelope" caused by the pixellation of the SLM, which would otherwise globally modulate the intensity pattern. Finally, in practice, we find that when we try to produce experimental patterns with $\mathcal{C}$ covering the maximum theoretically permitted area (25\% of the trapping plane), we cannot achieve patterns with less than 20\% RMS fluctuations. This is significantly improved in the results presented below by reducing $\mathcal{C}$ to cover only 10\% of the trapping plane.

\begin{figure}
\includegraphics[width=0.8\textwidth]{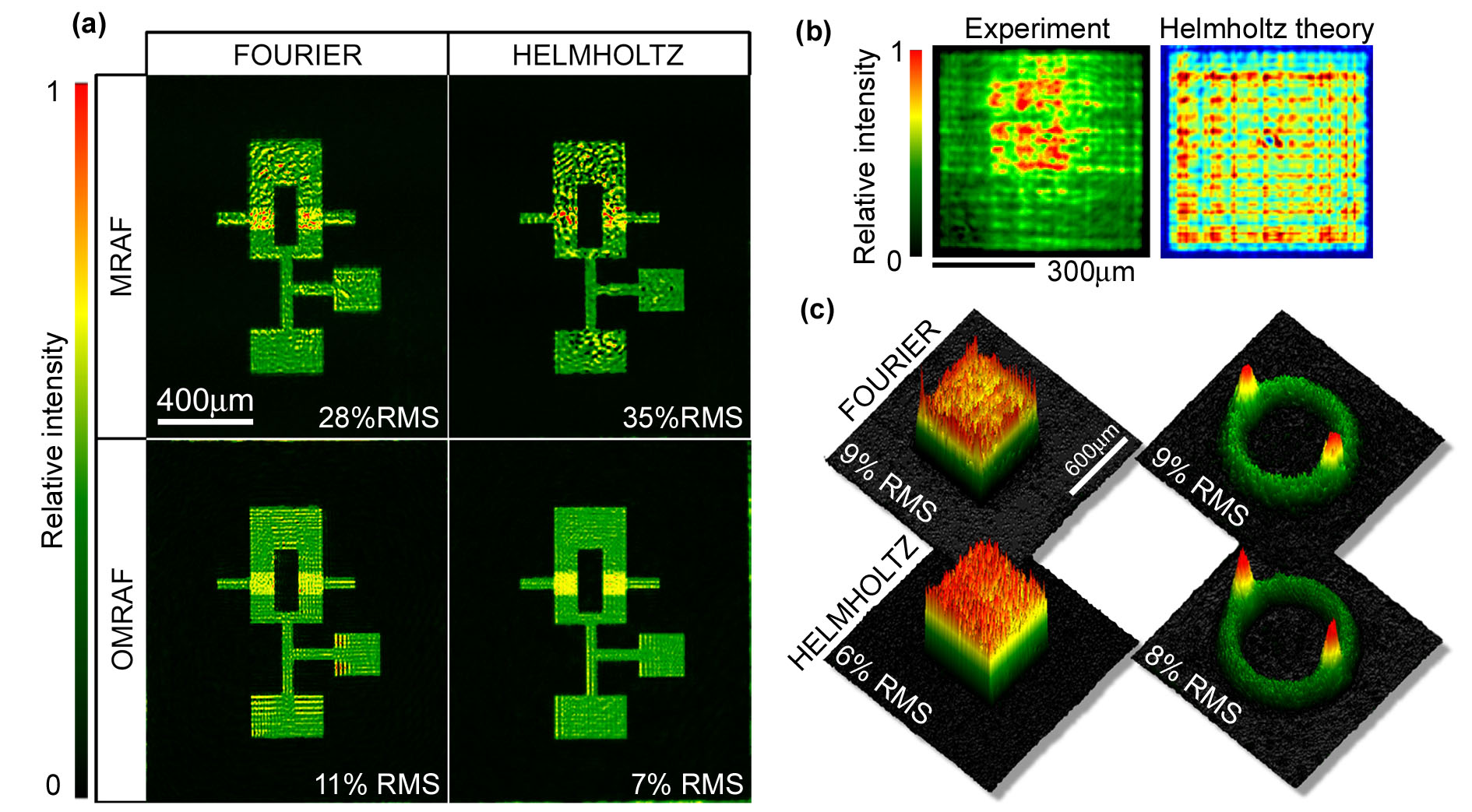}
\caption{Experimental results with 532nm laser light. (a) Experimental patterns based on four different computational methods (for display we crop the trapping plane around region $\mathcal{C}$). We observe low quality traps with the MRAF algorithm, and we are unable to improve them by adjusting the propagator (indeed, in this high RMS error regime, the additional complexity of the Helmholtz propagator can be detrimental). However, incorporating a Helmholtz propagator into our OMRAF algorithm allows us to produce the OR-gate with only 7\% RMS error. (b) Illustrates the need for a beyond-paraxial propagator. The left panel shows an experimental pattern based on OMRAF calculation under the paraxial approximation. The right panel shows that similar fringing is reproduced when the same paraxial kinoform is fed into a Helmholtz propagator to simulate the experiment. (c) To test the universality of our method, we also show experimental OMRAF patterns cropped to the boundary of $\mathcal{C}$ for a uniform square trap and an annular BEC stirrer.\label{fig:fig4}}
\end{figure}

In the previous section, we presented computational results generated under the paraxial approximation in which the propagator mapping the light in the SLM plane to the trapping plane is given by a scaled Fourier transform \cite{Ersoy07}. Using this approach in our experiments with an $f=200\,$mm lens, we are able to produce a trapping pattern with RMS error of $\approx11\%$  [Fig. \ref{fig:fig4}(a)]. As highlighted in Fig. \ref{fig:fig4}(b), the most prominent form of error is semi-regular fringing. We suggest that the source of this error is the inadequacy of the paraxial approximation, and use a numerical Helmholtz solver (see the Methods section) to test this hypothesis. Specifically, we use the OMRAF algorithm under the paraxial approximation to produce a ``Fourier-generated" kinoform, and then input this kinoform into our numerical Helmholtz propagator to simulate the experimental apparatus. This indeed produces the fringing pattern similar to the one observed experimentally, which gives us confidence to continue with the Helmholtz method. We thus replace all instances of the Fourier propagator in the OMRAF algorithm with a Helmholtz propagator.

In this way, we are able to eliminate the erroneous fringing, and reduce the RMS error of the experimental pattern to 7\% [see Fig. \ref{fig:fig4}(a)]. As shown in Fig. \ref{fig:fig4}(c), we can produce other trap shapes with similar fidelity. Thus, for traps with sharp edges and an extended canvas region, we achieve RMS variation comparable to the 4\% fluctuations previously seen only for simple smooth traps \cite{Bruce10}. In Ref. \cite{Wright12}, it has already been shown that 5\% RMS errors are sufficiently low for atomtronic applications. We therefore believe that our methods show great promise for future applications.

\section*{Discussion}
We have presented a simple method for producing optical traps of arbitrary 2D profile suitable for use in cold atom experiments. A useful optical potential must be referenced to a well defined background, and we showed that in order to do this, we must modify the existing algorithms to include a non-zero background light intensity. Specifically, we developed the OMRAF algorithm guided by the principle that points of zero intensity seed convergence-stalling vortices, and therefore should be avoided. In computer simulations we produce traps covering 25\% of the trapping plane (the Nyquist limit) with $4\%$ RMS error after only 30 iterations. We also took the next step of realising these 2D profiles in laser light. Here, we remove systematic aberrations using a Shack-Hartmann method and active feedback. In addition, we found that using a beyond-paraxial Helmholtz solver improves the fidelity of the experimentally produced light patterns. 

\section*{Methods}

\subsection*{Modified Gerchberg-Saxton Algorithms}
Both the MRAF and the OMRAF approach have the Gerchberg-Saxton algorithm at their core. This algorithm attempts to obtain the kinoform, $\phi(\mathbf{r})$, which must be applied to the input Gaussian field, $E_0(\mathbf{r})$, in the SLM plane (spanned by $\mathbf{r}$) to obtain the desired trap shape, $T(\mathbf{R})$, in the trapping plane (spanned by $\mathbf{R}$) at $z = 2f$ [see Fig. \ref{fig:fig1}(a)]. These two planes are related by a projection operator, $\mathscr{P}$, so we may write the implicit equation for $\phi(\mathbf{r})$ as:

\begin{figure}[t!]
\includegraphics[width = \textwidth]{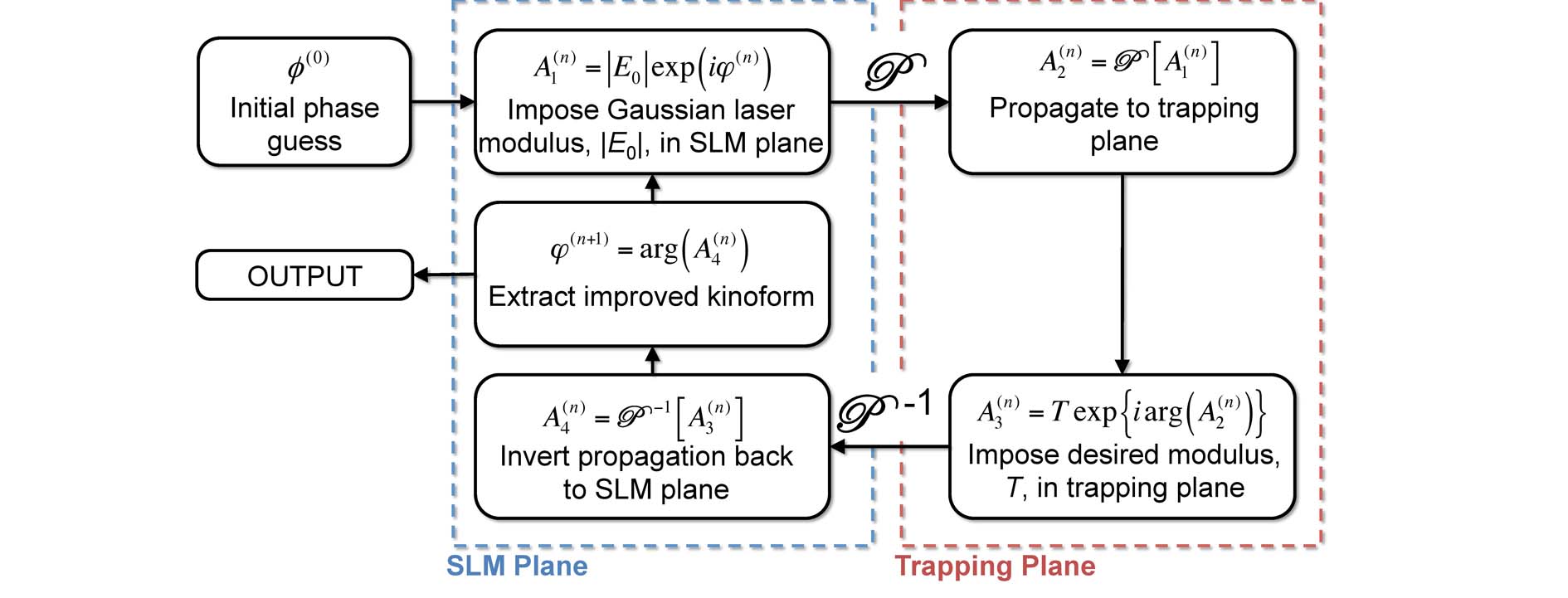}
\caption{Diagrammatic representation of the core Gerchberg-Saxton algorithm. Refer to the text for definition of symbols.\label{fig:fig5}}
\end{figure} 
\begin{equation}\label{eqn:PhaseRetrieval}
\left|\mathscr{P}\left[\left|E_0(\mathbf{r})\right|\exp\{i \phi(\mathbf{r})\} \right]\right|=T(\mathbf{R}).
\end{equation}
The Gerchberg-Saxton algorithm solves this equation iteratively as illustrated in Fig. \ref{fig:fig5}.

MRAF relaxes the modulus constraint outside the trapping region by defining a drawing, $\mathcal{D}$ (defined by $T(\mathbf{R})>0$) together with a narrow canvas, $\mathcal{C}$, of zero intensity around $\mathcal{D}$, and then applying the following algorithm:

\begin{equation}\label{eqn:MRAFoperator}
A_3^{(n)}(\mathbf{R})=\left\{ \begin{array}{l l} mT(\mathbf{R})\exp\left\{i\arg\left(A_2^{(n)}(\mathbf{R})\right)\right\} &\qquad \mathbf{R}\in\mathcal{D\cup C} \\  (1-m)A_2^{(n)}(\mathbf{R}) &\qquad \mathbf{R}\notin\mathcal{D}\cup \mathcal{C}\end{array}\right..
\end{equation}
We find that a mixing parameter $m=0.4$ gives the minimum RMS deviation for all the patterns trialled (see also \cite{Pasienski08}). 

OMRAF further modifies this algorithm in the following ways: First we expand the canvas so that the trapping region covers the maximum theoretical area set by Nyquist's theorem (25\% of the whole trapping plane - see the next section). Secondly, throughout $\mathcal{C}$ and $\mathcal{D}$, we offset the desired light intensity according to:

\begin{equation}
T(\mathbf{R})\to \sqrt{T(\mathbf{R})^2+\Delta^2} \;.
\end{equation}

This shifts both the trapping potential and its zero reference level equally, and does not change the physics of the trap.

\subsection*{Maximum Canvas Size}
In order to best use the degrees of freedom offered by an $N \times N$ pixel SLM, we should allow region $\mathcal{C}$ to cover as many pixels as possible in the discretized trapping plane. This theoretical limit is found by the following Nyquist argument: The $N \times N$ SLM plane can be fast-Fourier-transformed to an $N \times N$ trapping plane with a diffraction limited spot size of exactly 1 pixel. The Nyquist limit states that the diffraction limit should be reduced to \textit{half} a pixel (the smallest feature in the trap shape). This can be done by padding the computer-simulated SLM plane with zeros to increase its size to $2N \times 2N$. This means that only 25\% of the simulated SLM plane is actually covered by the physical SLM, so we can at best control only 25\% of the trapping plane.

\subsection*{A Computationally Efficient Helmholtz Solver}
We see improvement in the experimentally realised trapping intensities when the following method is used to model the propagation operator $\mathscr{P}$ in the OMRAF iterations:

The Helmholtz equation for light propagation in free space gives that the Fourier transform of a light field, $E(\mathbf{r};z)$, is modified by a phase factor $\Theta(\mathbf{\xi};z)$ as it propagates in the $+z$ direction \cite{Ersoy07}
\begin{equation}
\mathscr{F}\left[E(\mathbf{r};z); \mathbf{\xi}\right]=\mathscr{F}\left[E(\mathbf{r};0);\mathbf{\xi}\right]\underbrace{\exp\left\{ikz\sqrt{1-(\lambda\mathbf{\xi})^2}\right\}}_{\Theta(\mathbf{\xi};z)},
\end{equation}

\noindent where $\mathscr{F}\left[g(\mathbf{r});\mathbf{\xi}\right]$ denotes the Fourier transform (FT) of $g$, making the Fourier variable pairing $\mathbf{r}\leftrightarrow\mathbf{\xi}$ explicit, and $k=2\pi/\lambda$ denotes the wave vector of the light. The routine for propagation of $E_0$ via the apparatus in Fig. \ref{fig:fig1}(a) can then be conceptually decomposed into 3 stages:\\
\begin{center}
\renewcommand{\arraystretch}{1.5}
\begin{tabular}{r p{2in}}
\centering
``Propagate FT from SLM to lens" & $\psi_1(\mathbf{\xi})=\mathscr{F}\left[E_0(\mathbf{r});\mathbf{\xi}\right]\Theta\left(\mathbf{\xi};f\right)$\\
``Apply Lens in Fourier space" & $\psi_2(\mathbf{\eta}) = \left[\psi_1 \circledast \widetilde{L}\right](\eta)$ \\
``Propagate FT to trapping plane" & $\psi_3(\mathbf{\eta})=\psi_2(\mathbf{\eta})\Theta(\mathbf{\eta};f)$\\
\end{tabular}
\begin{equation}
\mbox{Then:} \qquad\mathscr{P}_{\rm Helmholtz}\left[E_0(\mathbf{r});\mathbf{R}\right]=\mathscr{F}^{-1}\left[\psi_3(\mathbf{\eta});\mathbf{R}\right],
\end{equation}
\end{center}

\noindent where $\circledast$ represents convolution, and $\widetilde{L}$ represents the FT of the phase pattern imprinted by the lens, which, after aberration correction, we assume to be:

\begin{equation}
\widetilde{L}(\mathbf{\xi})\approx\mathscr{F}\left[\exp\left(\frac{ik\mathbf{r}^2}{2f}\right); \mathbf{\xi}\right].
\end{equation}

\vspace{0.5cm}
\noindent \textbf{Acknowledgements} \hspace{0.5cm} We thank Naaman Tammuz, Stuart Moulder, Scott Beattie and Richard Bowman for helpful discussions, and Robert Smith for comments on the manuscript. This work was supported by EPSRC (Grant No. EP/G026823/1).

\vspace{0.5cm}
\noindent \textbf{Author Contributions} \hspace{0.5cm} AG performed the simulations and experiments. Both authors contributed equally to the conception of the project, discussion of the results and writing of the manuscript.

\vspace{0.5cm}
\noindent \textbf{Competing financial interests} \hspace{0.5cm} The authors declare no competing financial interests.

\end{document}